\documentclass[sigconf]{acmart}
\usepackage{tcolorbox}
\usepackage{enumitem}
\AtBeginDocument{%
  \providecommand\BibTeX{{%
    \normalfont B\kern-0.5em{\scshape i\kern-0.25em b}\kern-0.8em\TeX}}}

\setcopyright{acmlicensed}
\copyrightyear{2024}
\acmYear{2024}
\acmDOI{XXXXXXX.XXXXXXX}

\acmConference[Conference acronym 'XX]{Make sure to enter the correct
  conference title from your rights confirmation emai}{June 03--05,
  2018}{Woodstock, NY}
\acmISBN{978-1-4503-XXXX-X/18/06}

\begin{document}

\title{Large Language Models as Conversational Movie Recommenders: A User Study}

\author{Ruixuan Sun}
\affiliation{%
 \institution{Grouplens Research, University of Minnesota}
 \streetaddress{5-244 Keller Hall, 200 Union Street SE}
 \city{Minneapolis}
 \state{Minnesota}
 \country{United States}}

 \author{Xinyi Li}
\affiliation{%
 \institution{Northwestern University}
 \city{Evanston}
 \state{Illinois}
 \country{United States}}

 \author{Avinash Akella}
\affiliation{%
 \institution{Grouplens Research, University of Minnesota}
 \streetaddress{5-244 Keller Hall, 200 Union Street SE}
 \city{Minneapolis}
 \state{Minnesota}
 \country{United States}}

 \author{Joseph A. Konstan}
 \affiliation{%
 \institution{Grouplens Research, University of Minnesota}
 \streetaddress{5-244 Keller Hall, 200 Union Street SE}
 \city{Minneapolis}
 \state{Minnesota}
 \country{United States}}

\begin{abstract}

This paper explores the effectiveness of using large language models (LLMs) for personalized movie recommendations from users' perspectives in an online field experiment. Our study involves a combination of between-subject prompt and historic consumption assessments, along with within-subject recommendation scenario evaluations. By examining conversation and survey response data from 160 active users, we find that LLMs offer strong recommendation explainability but lack overall personalization, diversity, and user trust. Our results also indicate that different personalized prompting techniques do not significantly affect user-perceived recommendation quality, but the number of movies a user has watched plays a more significant role. Furthermore, LLMs show a greater ability to recommend lesser-known or niche movies. Through qualitative analysis, we identify key conversational patterns linked to positive and negative user interaction experiences and conclude that providing personal context and examples is crucial for obtaining high-quality recommendations from LLMs.
\end{abstract}


\maketitle

\section{Introduction}
Large language models (LLMs) have been developed extensively and applied to various domains to assist with a range of human tasks \cite{cascella2023evaluating, li2023large, belzner2023large}. Previous studies have examined the integration of LLMs into recommender systems (RecSys) using two main approaches: 1) Using LLMs in the RecSys training process, such as testing fine-tuning techniques for candidate generation \cite{zhang2021language}, generating latent space or item descriptions \cite{yang2023large, acharya2023llm, silva2024leveraging}, and designing personalized prompts for downstream recommendation tasks \cite{geng2022recommendation}; 2) Leveraging LLMs to simulate user agents \cite{huang2023recommender} or directly evaluate recommendation results \cite{di2023retrieval, xu2024prompting}.

However, there is a shortage of studies focusing on real user evaluations of LLM-based recommenders, an essential but challenging step in gauging practical recommender system (RecSys) performance \cite{konstan2012recommender}. Given their interactive and question-answering nature \cite{wu2022ai}, LLMs are ideal for tasks that aim to make recommendations more transparent to users \cite{bang2023examination}. Real user evaluations can offer deeper insights into recommendation quality by focusing on human-centric values like interpretability, trustworthiness, and why users like or dislike certain recommended items \cite{pu2012evaluating, knijnenburg2012explaining}. These evaluations can help clarify the underlying characteristics that contribute to user satisfaction with recommendations.

In this study, we explore the gap by understanding the quality of recommendations generated by pre-trained open-source LLMs under differently-personalized system prompts and three distinct scenarios. With hundreds of active users recruited from an online movie recommender system, we develop a chatbot user interface and assess their natural interaction experience to get personalized recommendations from LLMs in an online field experiment. By analysing survey response and users' conversation patterns, we address the following three research questions:

    \emph{\textbf{RQ1}: How do users perceive the LLM recommender compared to a classic recommender experience?}

   \emph{\textbf{RQ2}: How do different prompts, scenarios, or users' native consumption factors influence perceived LLM recommendation qualities?}

    \emph{\textbf{RQ3}: What are some effective interaction strategies users can employ with the LLM to attain more satisfactory recommendations?}

    \begin{figure*}[]
        \centering
        \includegraphics[width=\textwidth]{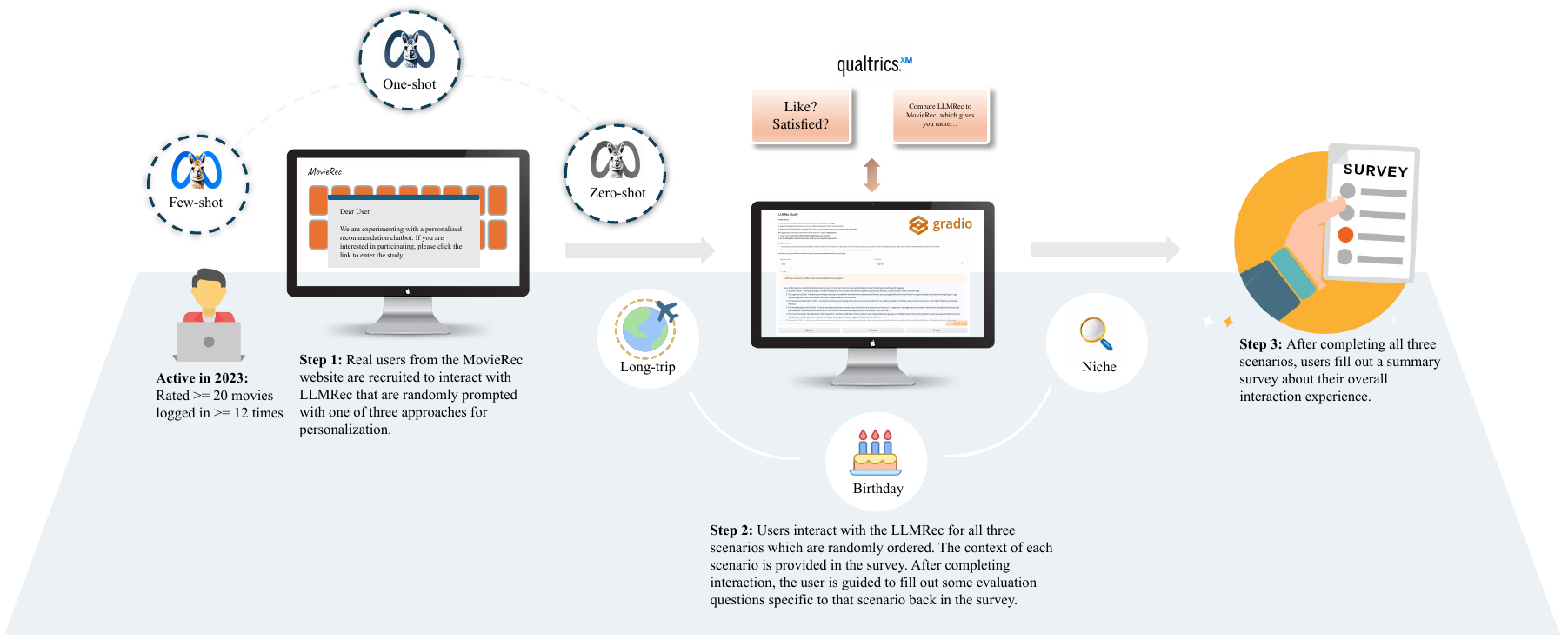}
        \caption{Overview of the LLM recommender user study design.}
        \label{fig:overall-study-diagram}
    \vspace*{-0.3cm}
    \end{figure*}

\textbf{We outline the main contributions of this paper as follows}:

\begin{itemize}[noitemsep,topsep=0pt]
    \item We confirm that out-of-the-box LLMs excel in providing explainable recommendations and creating an interactive user experience. However, they struggle with generating personalized, novel, diverse, serendipitous, and trustworthy movie recommendations compared to traditional recommenders. 
    \item We find that different prompting techniques do not significantly impact user-perceived recommendation quality and experience. We also reveal that user consumption history plays a significant role in how they appreciate recommendations generated by LLMs. Moreover, LLMs better accommodate niche or unpopular movie recommendations, compared to more personalized or ask-for-others requests. 
    \item An analysis of user conversation patterns shows that providing context or examples is crucial for improving recommendation quality and user satisfaction with LLM recommenders. 
\end{itemize}

In the rest of this paper, we first review relevant literature, then share details of the study procedure, followed by presenting the results of RQs and analysis with both statistical and qualitative data from the survey response. Finally, we synthesize the findings and design implications for future LLM-RecSys applications. 

\section{Related Work}

\subsection{Large Language Models in Recommendation}

Trained on vast textual corpora, LLMs have demonstrated remarkable capabilities on a variety of tasks relying on textual instructions \cite{radford2019language}. Their applications extend beyond natural language processing (NLP) \cite{qin2023chatgpt} into fields like education \cite{baidoo2023education, kasneci2023chatgpt} and medicine \cite{thirunavukarasu2023large}. To harness the full potential of LLMs, several studies have examined prompting techniques to improve their performance. These techniques include one-shot or few-shot prompting, where the model learns from provided examples \cite{logan2021cutting}, and more advanced strategies such as `Chain of Thought' (CoT) \cite{wei2022chain} and `Tree of Thoughts' (ToT) \cite{yao2024tree}, which introduce structured reasoning to generate responses.  

As LLMs' capabilities grow, an increasing body of research focuses on their use in RecSys. This work often frames recommendation tasks as language-based problems and adapts LLM architectures \cite{Li2023Explain, li2024prompt, geng2022recommendation, Zhang_2023}. Another emerging trend is applying LLMs to RecSys tasks in a zero-shot setting. For example, \citeauthor{liu2023chatgpt} and \citeauthor{dai2023uncovering} examined ChatGPT's potential for RecSys tasks \cite{liu2023chatgpt, dai2023uncovering}, while \citeauthor{he2023} explored LLMs as zero-shot recommenders in conversational settings \cite{he2023}. \citeauthor{sanner2023large} found that LLMs are competitive with cold-start recommenders, especially with natural language preferences, and that few-shot prompting outperforms zero-shot \cite{sanner2023large}. However, these studies mainly focused on offline evaluations, using LLMs solely as recommendation engines while neglecting online user experiences.

To bridge the gap between theoretical assessments and real-world applications of LLM-based recommenders, we conduct live experiments with actual users of an operating movie recommender site. These experiments involve users interacting with LLMs via a real-time chatbot interface to get recommendations. Our study examines not only the impact of prompting techniques on offline RecSys tasks, as outlined in \cite{Liu2023IsCA}, but also how different prompting approaches—specifically zero-shot, one-shot, and few-shot—affect user experiences in practical recommendation scenarios. By gathering direct feedback from users, we gain valuable insights into their engagement and evaluate the LLM's recommendation performance in real-world conditions.
 
\subsection{RecSys User Experience}

User experience studies in recommendation systems (RecSys) are crucial for creating solutions that meet user needs and expectations. Researchers like \citeauthor{pu2011user} and \citeauthor{knijnenburg2012explaining} proposed human-centric frameworks to evaluate RecSys user experiences \cite{pu2011user, knijnenburg2012explaining}. \citeauthor{agner2020recommendation} examined the effectiveness of machine-learning-based streaming media recommendation systems, focusing on users' mental models \cite{agner2020recommendation}. \citeauthor{liu2024evaluating} investigated how different types of explanations could improve user trust and satisfaction through online experiments \cite{liu2024evaluating}. \citeauthor{narducci2018improving} explored user interactions with chatbot recommenders powered by PageRank algorithms \cite{narducci2018improving}. However, these studies did not specifically focus on user experiences when interacting with LLMs in RecSys.

\citeauthor{granada2023videolandgpt} conducted a study on an LLM-based conversational RecSys, assessing its performance with personalized and non-personalized recommendations. They evaluated accuracy, safety, and fairness \cite{granada2023videolandgpt}. \citeauthor{silva2024leveraging} leveraged ChatGPT to generate personalized movie recommendation explanation to users and evaluated user perception of personalization, effectiveness, and persuasiveness with a user study \cite{silva2024leveraging}. In contrast, our study focuses on user experiences with recommendations generated directly by the LLM, without relying on predefined candidate lists. Along with evaluating the impact of personalized and non-personalized recommendations, we also examine how different prompting techniques affect user experiences. Using metrics like diversity, novelty, serendipity, trustworthiness, and explainability, we ask users to compare recommendations from a baseline RecSys with those from an LLM-based system. This helps understand how to best integrate LLMs into existing RecSys to improve user experiences.

\section{Study Design}

    \begin{figure*}[]
        \centering
    \includegraphics[width=\textwidth]{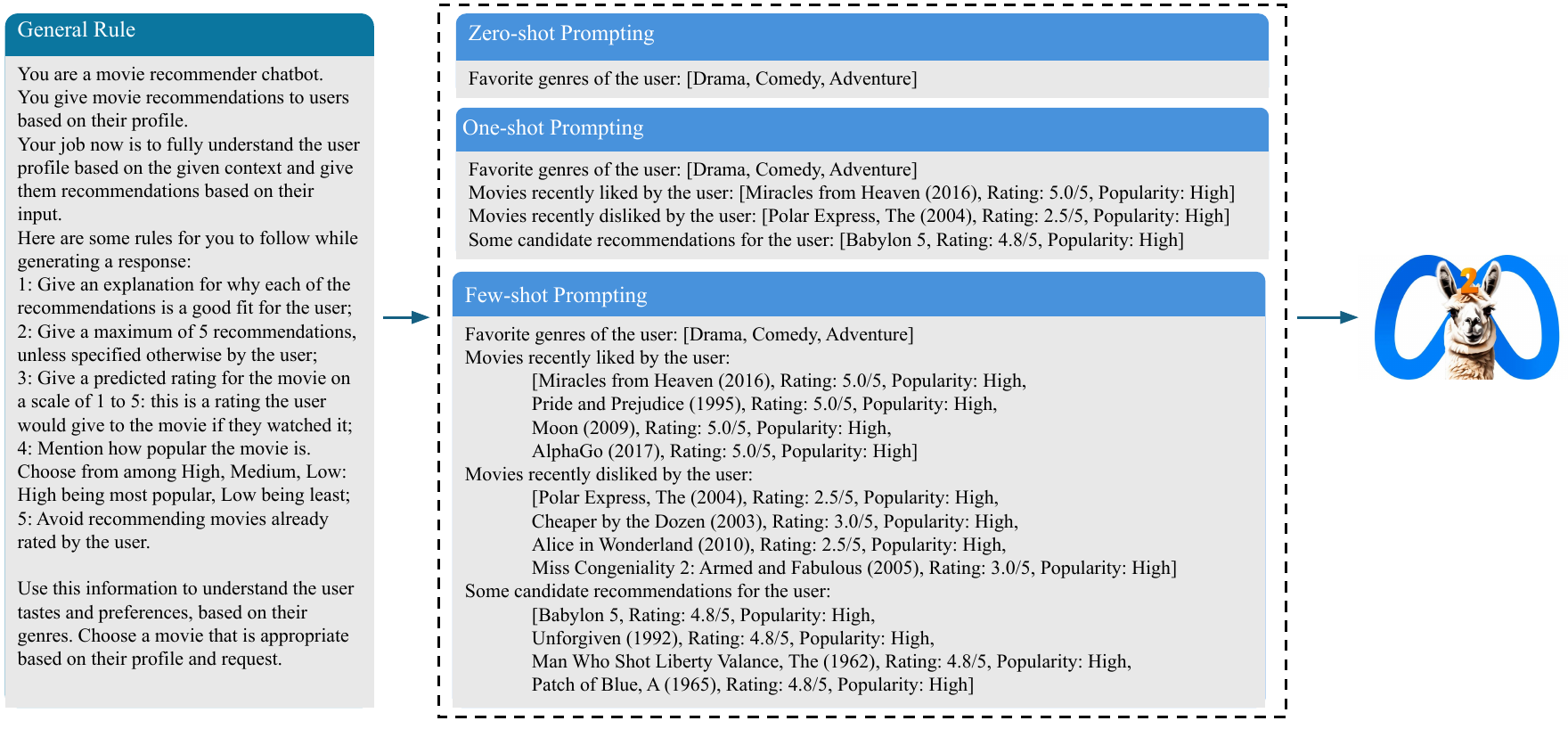}
    \caption{Diagram depicting the general rule and randomly selected prompting technique of the LLM for engaging with users at the 1st phrase of the user study.}
    \label{fig:system-prompt}
    \vspace{-0.3cm}
    \end{figure*}

 As illustrated in Fig. \ref{fig:overall-study-diagram}, our study design can be split into three phases: 1) participant recruitment and personalized system prompt generation; 2) interaction with LLM-based recommender under different scenarios; and 3) scenario-based and summary surveys including various evaluation metrics.

\subsection{Participants and Personalized Prompts} 

We partnered with a movie recommendation platform (\textbf{MovieRec}\footnote{The platform's real name is anonymized and will be revealed upon publication of this paper.}) to select qualified participants for our study. To ensure that users had substantial experience with recommender systems, providing a baseline for comparison in survey questions, we recruited only active users who logged into the system at least 12 times and rated more than 20 movies in 2023. Using these criteria, we identified 3,031 qualified users from the platform's database. Recruitment messages and access to the questionnaire were displayed prominently on users' homepages upon their initial login during the experiment period. Participation was entirely voluntary, with no incentives provided, promoting more thoughtful responses to survey questions. Our study design was determined as a non-human subject study by our institutional review board (IRB).

For the recommendation task, we used the pre-trained Llama2-7b-Chat \cite{touvron2023llama}, an open-source model released by Meta in 2023, designed for conversational dialogues \footnote{Download link: https://huggingface.co/meta-llama/Llama-2-7b-chat-hf}. Due to privacy concerns related to user data, we hosted the model on a local server instead of using a cloud-based API. We differentiated three prompting techniques based on the level of personalized user context provided, as outlined in Fig. \ref{fig:system-prompt}. Beyond the shared general guidelines, the zero-shot prompt only contained the top three rated genres based on each user's historical ratings. The one-shot prompt included three additional lists that each contains one movie most recently liked (top 10\% of user's all ratings), disliked (bottom 10\% of user's all ratings), and recommended for the user based on the platform's existing recommendation model. The few-shot prompt extended this by listing four movies in each category. This design was inspired by previous research, suggesting that providing more than five example movies could lead to unstable outcomes \cite{sileo2022zero}. Additionally, we categorized each movie's popularity into three tiers: high, medium, and low, based on their respective percentiles in the MovieRec database based on its total number of user ratings (high: top 20\%, medium: 60\%-80\%, low: below 60\%).

\vspace{-0.1cm}
\subsection{User Interface and Scenarios}

For user interaction, we developed the chatbot interface using Gradio \footnote{https://www.gradio.app/}, as shown in Appendix Fig. \ref{fig::example_chatbot_ui}. Throughout the rest of the paper, we'll refer to this interface as \textbf{LLMRec}. Users were encouraged to interact with the chatbot for as long as they needed in each scenario. To mitigate potential latency issues arising from multiple users making inference requests, we ran pilot tests to ensure the model could handle real-time inference for at least two users at once. Additionally, we included a disclaimer message and an estimated wait time on the chatbot interface to keep users informed about potential generation queues during their interactions.

\begin{figure}[!htp]
        \centering
        \includegraphics[width=0.8\columnwidth]{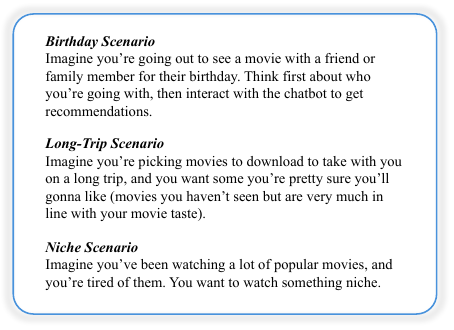}
        \caption{Three recommendation scenarios.}
        \label{fig:scenarios}
    \vspace*{-0.5cm}
    \end{figure}

Upon entering the survey, users were greeted with an informational page followed by three common movie recommendation scenarios: 1) \textit{Ask-for-others -- Friend's or Family's Birthday}; 2) \textit{Personalization -- Long Trip}; and 3) \textit{Unpopular -- Something Niche}. Detailed description of each scenario is included in Fig. \ref{fig:scenarios}. To mitigate potential position bias, the order of scenarios was randomized for each user within the questionnaire. This ensured that users' prior experiences did not influence their responses to subsequent scenarios.

\vspace{-0.2cm}
\subsection{Evaluation Metrics}

\begin{figure*}[!htp]
        \centering
        \includegraphics[width=2.0\columnwidth]{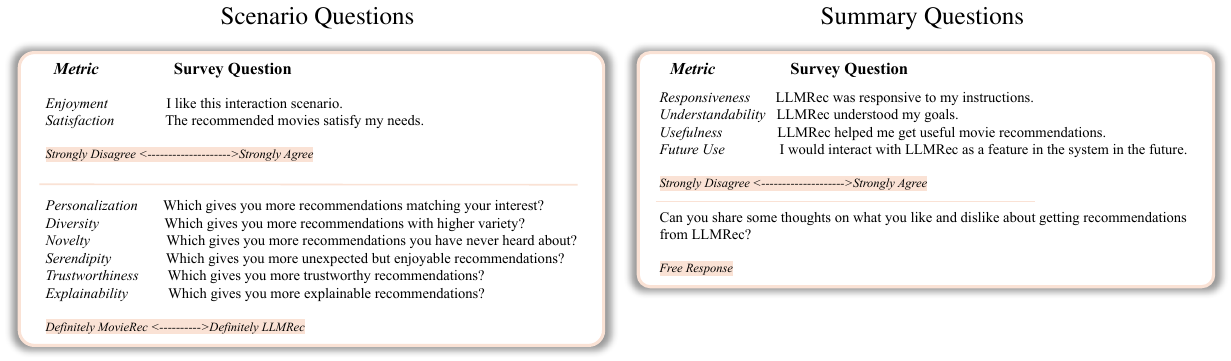}
        \vspace{-0.2cm}
        \caption{Survey Question as evaluation metrics.}
        \label{fig:metrics}
    \vspace*{-0.3cm}
    \end{figure*}

When finishing their conversation with LLMRec in each scenario, users were directed back to the survey to answer a set of questions. These questions covered: 1) their general appreciation of the recommendation scenario, focusing on enjoyment and satisfaction; and 2) a comparison between their experiences with MovieRec and LLMRec, measured by factors such as \textit{Personalization, Diversity, Novelty, Serendipity, Trustworthiness}, and \textit{Explainability}. The detailed phrasing of the survey questions for each scenario is shown on the left side of Fig. \ref{fig:metrics}, inspired by previous designs from \citeauthor{ekstrand2014user} and \citeauthor{nguyen2016enhancing} \cite{nguyen2016enhancing, ekstrand2014user}. 

After completing the scenario evaluations, users were directed to a summary page to provide feedback on their overall interaction experience with LLMRec. Drawing inspiration from chatbot survey designs in \citeauthor{chaves2021should}'s work, we asked users to rate various aspects of chatbot interaction, including \textit{Responsiveness, Understandability, Helpfulness}, and whether they would like to see LLMRec integrated into the MovieRec platform in the future \cite{chaves2021should}. The survey concluded with an optional free-response question, allowing users to share additional, unstructured insights.

\vspace{-0.2cm}
\section{Results}

    Our experiment ran for one month from 02/13/2024 until 03/14/2024. In total, 449 users enrolled in the study (i.e. opening the questionnaire), 178 submitted the questionnaire, and we build our analysis with 160 unique users who have completed all quantitative questions except the optional free response. The distribution of prompting techniques is relatively equal in size, with 56 users in zero-shot, 48 users in one-shot, and 56 users in few-shot bucket. The median of user response time is 17.62 minutes for completing evaluation of all three scenarios.

\vspace{-0.2cm}
\subsection{User-assessed LLMRec Quality}

\begin{figure}[!htp]
        \centering
        \includegraphics[width=\columnwidth]{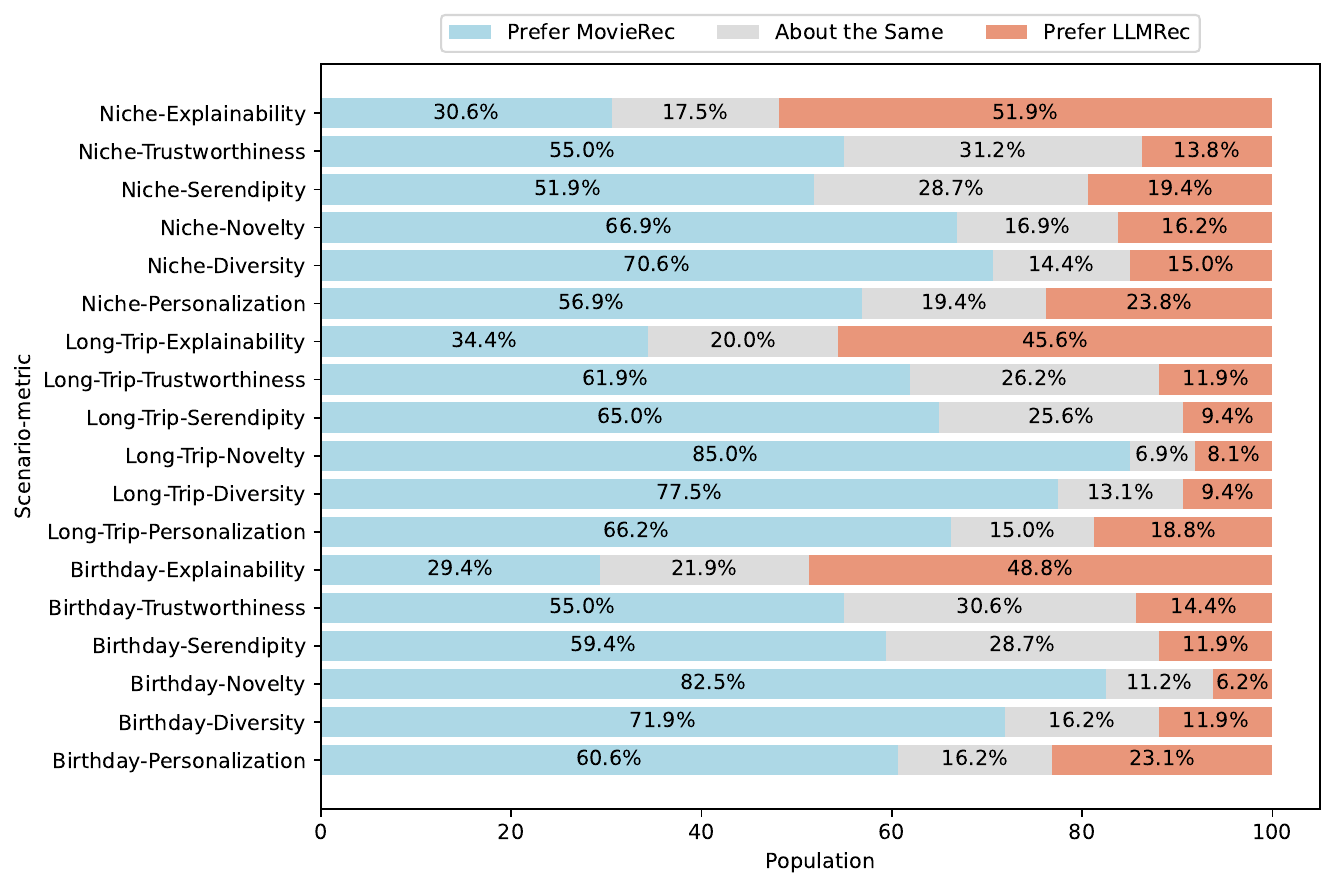}
        \vspace{-0.5cm}
        \caption{User perception of recommendation quality from LLMRec compared to the classic MovieRec experience. All LLMRec-MovieRec pairs are tested with paired t-test and shows p <= 0.05 statistical significance except for the Birthday-Explainability and Long-Trip-Explainability.}
        \label{fig::ml-chatbot-perf}
    \vspace*{-0.7cm}
    \end{figure}

Based on user survey responses comparing their MovieRec to LLMRec experiences, we categorized responses into three groups: \textit{prefer MovieRec}, \textit{About the Same}, and \textit{prefer LLMRec}. In comparison to MovieRec, LLMRec stood out for its explainability, but users generally found MovieRec to offer more novel, diverse, accurate, and trustworthy recommendations across all three scenarios (Fig. \ref{fig::ml-chatbot-perf}).

To gain a deeper understanding of individual user experiences, we conducted a qualitative analysis of the free-response data (N=137) using the classic grounded theory method (GMT) \cite{bryant2013grounded}. In this approach, three researchers who were trained in GMT independently coded the user responses. To ensure accuracy, we cross-checked 10\% of each other's coding at random. Subsequently, all researchers met to do thematic analysis \cite{clarke2017thematic}, assigning each code to one or more topic clusters and iterating this process until all ambiguities were resolved. Ultimately, we identified three major themes.

\subsubsection{Defect of Recommendation Quality} The first major theme we identified revolves around the attributes where LLMRec fell short. Many users complained about a lack of personalization (N=39) and novelty (N=56) in the recommendations they received. As participant P50 remarked, \emph{"...I received repetitive recommendations for movies I had already watched. While it correctly grasped the concept of gore and violent films, it predominantly suggested love stories for me and my wife, which lacked the variety I anticipated."} Another significant issue users noted was low diversity (N=10) and high-popularity (N=18). P376 summarized this by saying, \emph{"It pretty much always stuck to only the most popular movies of all time."}

\subsubsection{Interactivity, Explainability, and Context} In this theme, users shared their positive interaction experience with LLMRec (N=49). P153 praised, \emph{"I like that it is interactive and takes many of my already established preferences into consideration when making recommendations."} Building on this interactivity, users like P368 also enjoyed getting explanations on why they got certain recommendations, \emph{"I like that it explains why I might like the films--this is far better than the traditional MovieRec interface."} In addition to explanations, users found LLMRec supported context-oriented recommendation well, accommodating mood, interests of others, or even quickly extracting preferences from new users. As P407 remarked,\emph{"I like the additional option to tell the chatbot some information about what I want to watch which I can't do if I only rate movies. Additionally it can be helpful to get recommendations faster, if you're new to recommender services and haven't rated that much yet."}

\vspace{-0.3cm}
\subsubsection{Control, Trust, and Transparency} 

Despite generally positive interactions, users expressed concerns about the reliability of LLMRec (N=42). One issue was the difficulty in user control. As P169 shared, \emph{"...In the road trip scenario, it got fixated on horror movies and started only suggesting those, even though I dislike them very much."} Similar to previous empirical study on LLMs \cite{huang2023survey}, we also observed hallucination in recommendation scenarios, which deeply undermines user trust. P47 commented, \emph{"I am also wary about the chatbot suggestions because I know it can often hallucinate and not understand me clearly etc. I feel like I can't rely fully on chatbot, compared to relying on my own judgement and judgement of other users that in regular movie selection experience."} This sense of disappointment prompted users to call for more transparency in LLM recommendations. P23 noted, \emph{"It was also a black box, I like knowing that the algorithm is SVD based (for example), as opposed to whatever the collaborative intelligence model has scraped from the internet."} With data from the MovieRec-LLMRec comparison and qualitative analysis, we answer \emph{\textbf{RQ1}}:

    \emph{\textbf{RQ1}: How do users perceive the LLM recommender compared to a classic recommender experience?}

Our findings suggest that while users found LLM recommenders superior to classic systems in terms of explaining recommended movies, they also noted a lack of algorithmic transparency in the recommendation process. Users appreciated the LLM's smooth interaction flow and its ability to adapt to customized recommendation contexts. However, the recommendations were often non-personalized and skewed toward popular, homogeneous content, which diminished user trust and reduced the perceived reliability of the LLM recommender.

\subsection{Effect of Prompt, Scenario, and Rating}

\begin{figure}[!htp]
        \centering
        \includegraphics[width=0.8\columnwidth]{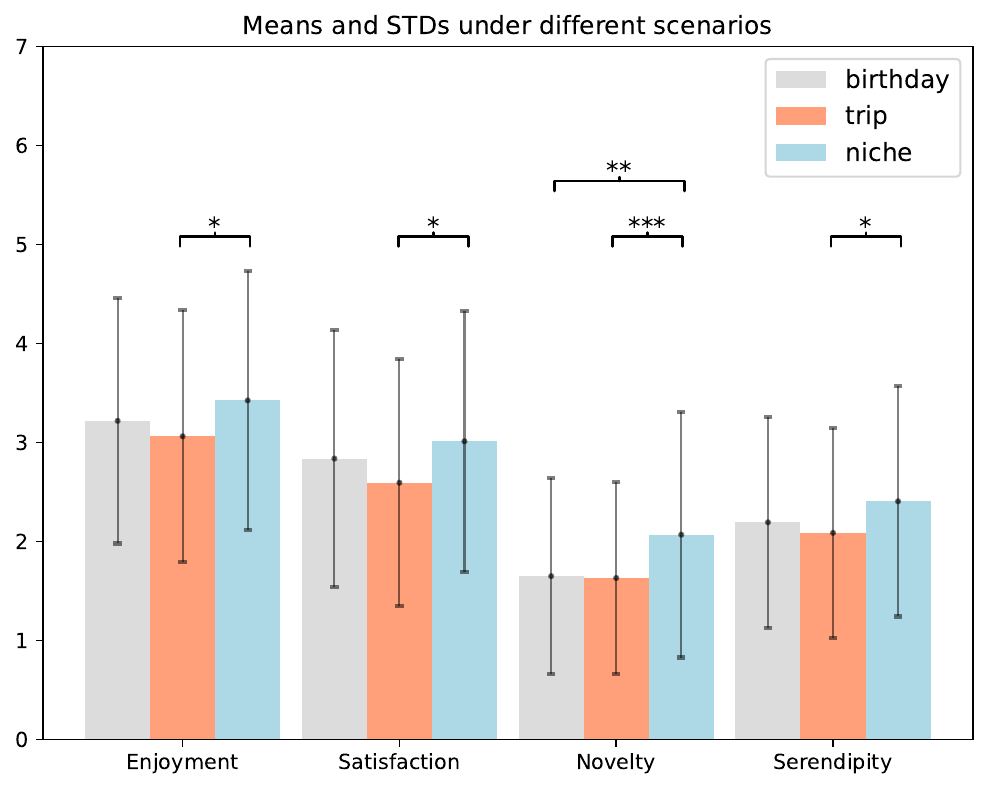}
        \includegraphics[width=\columnwidth]{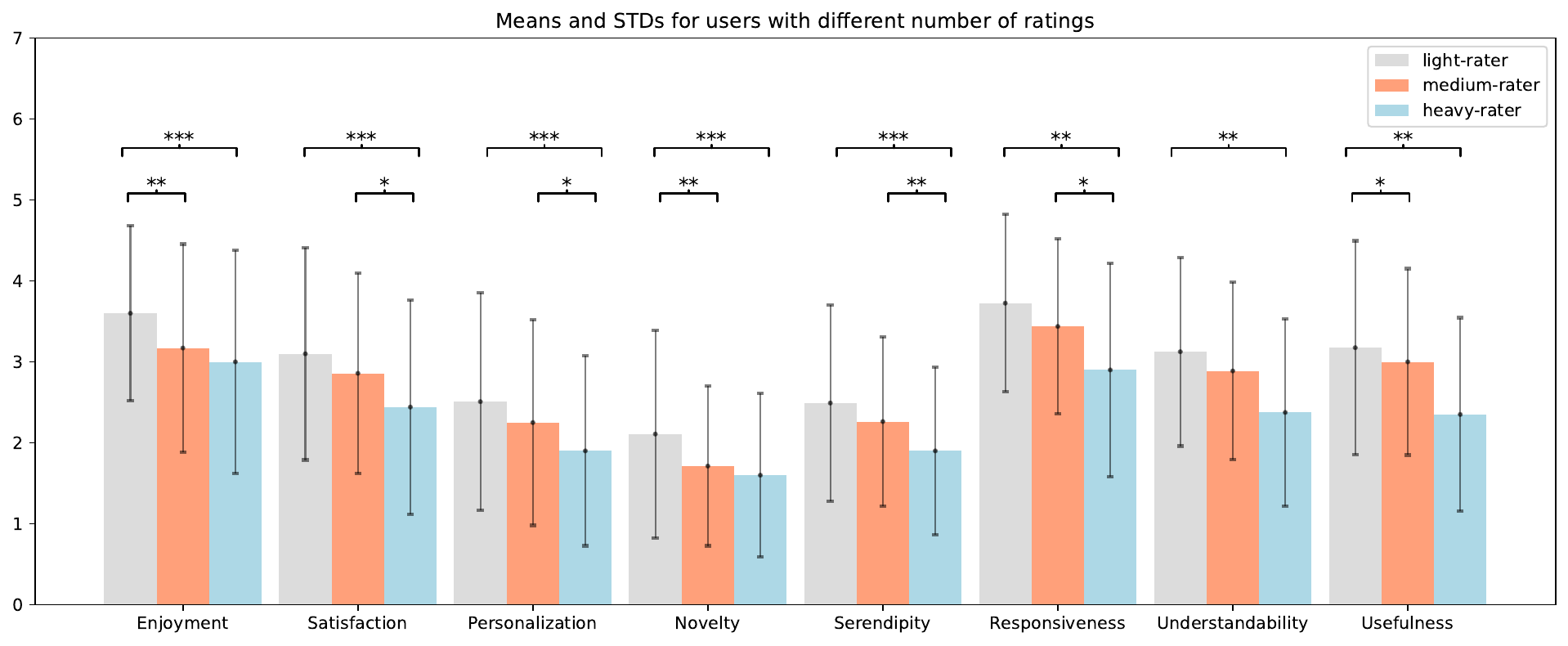}
        \caption{Within-subject and between-subject test results. Top: Differences of recommendation quality from different scenarios. Bottom: Differences of recommendation quality for users based on their historic rating count. Asterisk (*) indicates p-val between conditions with Tukey's HSD test:  $p<$ .1 (*); $p<$ .05 (**);  $p<$ .01 (***).}
        \label{fig:scenario_rating_stats}
    \vspace*{-0.6cm}
    \end{figure}

In the second stage of our analysis, we used a between-subject prompt-wise test and a within-subject scenario-wise test to examine differences across various response groups. We applied one-way ANOVA and pairwise Tukey's HSD \cite{abdi2010tukey} tests to assess statistical significance. Surprisingly, we did not find significant differences in evaluation metrics between different prompts (see Appendix Table \ref{tab:prompt-stats}). However, our ANOVA tests revealed group-wise statistical differences in the means for \emph{Enjoyment, Satisfaction, Novelty}, and \emph{Serendipity} scores (Fig. \ref{fig:scenario_rating_stats} and Table \ref{tab:scenario-stats}). Notably, users found the \emph{Niche} scenario more enjoyable and satisfying compared to the \emph{Trip} scenario. They also felt they received more novel and serendipitous recommendations in the \emph{Niche} scenario than in the other two scenarios.

    \begin{table}[h]
    \begin{tabular}{@{}llllll@{}}
    \toprule
                          & \textbf{F-val} & \textbf{P-val} & \textbf{B-T} & \textbf{B-N} & \textbf{T-N} \\ \midrule
    \textit{Enjoyment}    & 3.244          & .040           & -            & -            & .363 (.031)   \\
    \textit{Satisfaction} & 4.239          & .015           & -            & -            & .419 (.011)   \\
    \textit{Novelty}      & 8.461          & .000           & -            & .410 (.002)   & .438 (.001)   \\
    \textit{Serendipity}  & 3.476          & .032           & -            & -            & .319 (.027)   \\ \bottomrule
    \end{tabular}
    \caption{AVOVA and pairwise Tukey's HSD test results for different scenario metrics. Pairwise stats include the mean difference and p-val in bracket. N: Niche, T: Trip, B: Birthday.}
    \label{tab:scenario-stats}
    \vspace*{-0.8cm}
    \end{table}

The absence of differences across prompting techniques led us to investigate other potential factors affecting user perceptions beyond LLM attributes. By querying the MovieRec database, we categorized users into three groups based on their total number of ratings, as outlined in Table \ref{tab:rating-buckets}. This allowed us to determine if a user's historical movie consumption correlated with variations in recommendation and chatbot quality metrics (Fig. \ref{fig:scenario_rating_stats} and Table \ref{tab:rater-stats}).

Using ANOVA, we found statistically significant differences among these groups for several metrics, including recommendation \emph{Enjoyment, Satisfaction, Personalization, Novelty, Serendipity}, and chatbot \emph{Responsiveness, Understandability}, and \emph{Usefulness}. Light raters found LLMRec more enjoyable and novel compared to medium and heavy raters. Heavy raters, however, were less satisfied with LLMRec, reporting it was less personalized, less responsive to their requests, and less likely to generate unexpected recommendations. They also indicated that LLMRec did not understand their needs as well as it did for those who had rated fewer movies in the past.

    \begin{table}[]
    \begin{tabular}{@{}llll@{}}
    \toprule
    \textit{\textbf{}}    & \textbf{percentile} & \textbf{\# ratings} & \textbf{\# users} \\ \midrule
    \textit{light-rater}  & \textless{}=25\%    & \textless{}=406     & 40                \\
    \textit{medium-rater} & 25\ -75\%         & 406-1522            & 80                \\
    \textit{heavy-rater}  & \textgreater{}=75\% & \textgreater{}=1522 & 40                \\ \bottomrule
    \end{tabular}
    \caption{User rating buckets}
    \label{tab:rating-buckets}
    \vspace*{-1.0cm}
    \end{table}

    \begin{table}[]
    \resizebox{\columnwidth}{!}{%
    \begin{tabular}{@{}llllll@{}}
    \toprule
                               & \textbf{F-val} & \textbf{P-val} & \textbf{L-M} & \textbf{M-H} & \textbf{L-H} \\ \midrule
    \textit{Enjoyment}         & 7.360          & .001           & -.429(.007)  & -            & .600(.001)   \\
    \textit{Satisfaction}      & 8.195          & .000           & -            & -.417(.011)  & -.658(.000)  \\
    \textit{Personalization}   & 6.976          & .001           & -            & -.350(.037)  & -.609(.001)  \\
    \textit{Novelty}           & 7.718          & .001           & -.396(.003)  & -            & .508(.001)   \\
    \textit{Serendipity}       & 9.087          & .000           & -            & -.363(.009)  & .592(.000)   \\ \midrule
    \textit{Responsiveness}    & 5.290          & 006            & -            & -.538(.047)  & -.825(.005)  \\
    \textit{Understandability} & 4.626          & .011           & -            & -            & -.750(.010)  \\
    \textit{Usefulness}        & 5.349          & .006           & -            & -.650(.018)  & -.825(.008)  \\ \bottomrule
    \end{tabular}%
    }
    \caption{AVOVA and pairwise Tukey's HSD test results for different rater metrics. Pairwise stats include the mean difference and p-val in bracket. L: Light, M: Medium, H: Heavy.}
    \label{tab:rater-stats}
    \vspace*{-0.9cm}
    \end{table}

    We then answer RQ2 with our findings:
    
   \emph{\textbf{RQ2}: How do different prompts, scenarios, or users' native consumption factors influence perceived LLM recommendation qualities?}

We found no significant differences in user perceptions based on varying personalization prompts. Among the three scenarios, users derived more enjoyment and felt their needs were better met when asking for niche recommendations. This scenario also provided users with more novel and unexpected movie suggestions. Additionally, the level of users' past movie consumption proved to be a critical factor influencing their satisfaction with LLM recommendations and the chatbot's overall performance. Users who had rated more movies tended to find the recommendations less satisfactory and personalized, which likely contributed to a more negative view of the LLM recommender's responsiveness and overall usefulness.

\subsection{Useful Conversation Strategies}

In our third analysis stage, we examined user conversation data with LLMRec to identify relationships and patterns tied to different types of user experiences. Our initial step involved running an Ordinal Linear Regression (OLR) to evaluate the correlation between the number of sentences users exchanged with LLMRec and their satisfaction and understandability score. The results indicated significant negative relationships for both Satisfaction ($coef$ = -0.079, $p$ = 0.041) and Understandability ($coef$ = -0.118, $p$ = 0.002), implying that users who engaged in shorter dialogue feel the LLMRec had a better overall goal understanding. Also, those who engaged in longer conversations with LLMRec felt were less likely to be satisfied with their recommendations. 

\subsubsection{Pattern Coding}

Building on this initial finding, we conducted a qualitative analysis to identify conversation patterns associated with user responses to the \emph{Satisfaction} question. We categorized responses into positive (i.e., "Agree" and "Strongly agree") and negative (i.e., "Disagree" and "Strongly disagree"). Using the GMT and thematic analysis, we identified 10 key patterns that correlated with both positive and negative user conversations.

    \begin{figure}[!htp]
      \vspace*{-0.2cm}
        \centering
        \includegraphics[width=0.8\columnwidth]{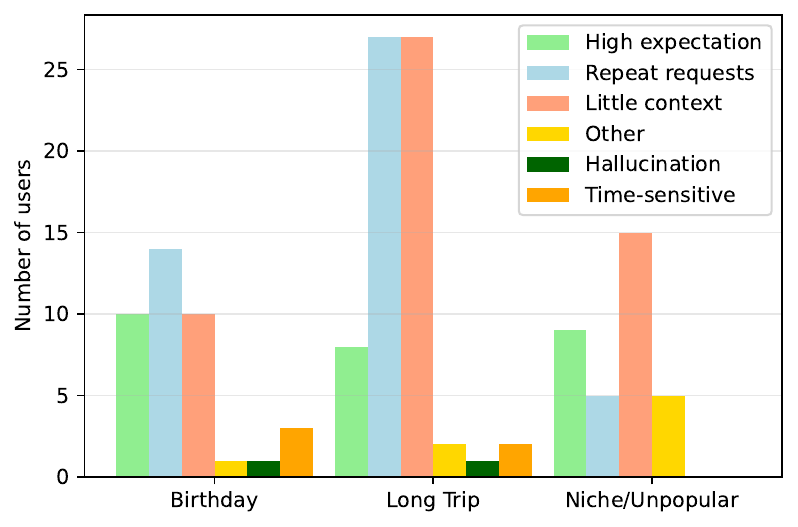}
        \includegraphics[width=0.8\columnwidth]{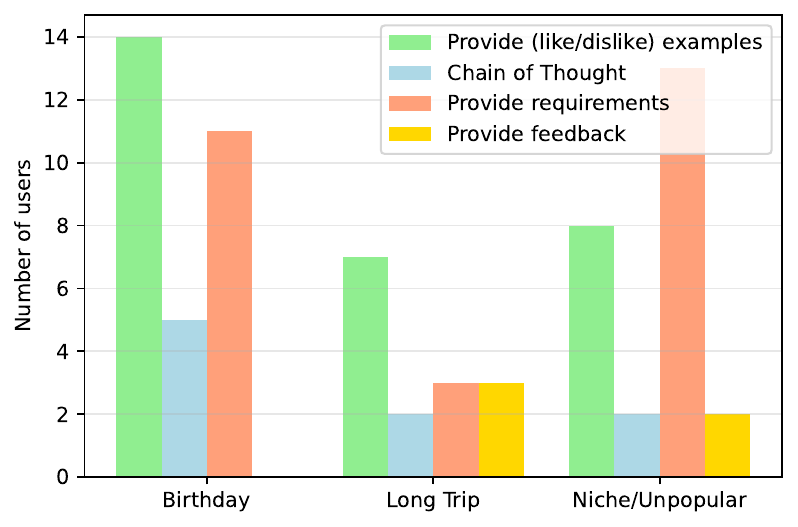}
        \vspace*{-0.2cm}
        \caption{User conversation patterns based on qualitative coding. Top: the distribution of patterns in negative conversation. Bottom: the distribution of patterns in positive conversation.}
        \label{fig:convo_coding}
    \vspace*{-0.4cm}
    \end{figure}

Figure \ref{fig:convo_coding} illustrates that when users provided specific information such as previously watched movies, expressed their preferences (likes or dislikes), or specified genres, their satisfaction with recommendations tended to be higher. In contrast, users who offered minimal context and expected LLMRec to infer all their preferences on its own, treating it like a search engine, generally had lower satisfaction. Based on these insights, we refined the conversation patterns and defined five discrete tags to characterize conversation attributes, as shown in Table \ref{tab:convo_tags}. Some example conversation classified under these tags can be found in Appendix Table \ref{tab:example-user-convo-tagging}.

    \begin{table*}[]
    \resizebox{2\columnwidth}{!}{%
    \begin{tabular}{@{}ll@{}}
    \toprule
    \textbf{Tag} &
      \textbf{Description} \\ \midrule
    \textit{Dialogue} &
      \begin{tabular}[c]{@{}l@{}}When users have a conversation by sharing details and giving feedback on how they feel about the recommendations. It should \\ also be like talking to humans instead of a search engine, not always dumping questions or steering the response to what they want.\end{tabular} \\ \cmidrule(l){2-2} 
    \textit{Context} &
      \begin{tabular}[c]{@{}l@{}}When users clearly specify their tastes/requirements/preferences and what they're looking for.  Specifically, they need to come up \\ with something on their own, not just reuse the scenario context we provided in the survey.\end{tabular} \\ \cmidrule(l){2-2} 
    \textit{Steering} &
      \begin{tabular}[c]{@{}l@{}}When users try to make the recommendations converge to their choice without clarifying their objective. They may have multiple\\ rounds of conversation and iteratively try to nudge the chatbot to get closer to their goal.\end{tabular} \\ \cmidrule(l){2-2} 
    \textit{Testing} &
      \begin{tabular}[c]{@{}l@{}}When users try to see how the system works and usually test it negatively. Some examples include asking questions outside of \\ movie recommendations, or trying to decode the system prompts the bot was based on, or using inappropriate language.\end{tabular} \\ \cmidrule(l){2-2} 
    \textit{Retry} &
      When users re-generate a response to their previous query. This can be identified by repeated queries in conversation history. \\ \bottomrule
    \end{tabular}%
    }
    \caption{Main tags associated with user conversation with LLMRec.}
    \label{tab:convo_tags}
    \vspace*{-0.8cm}
    \end{table*}

\subsubsection{Regression Model}

We further analyzed all 417 user conversation queries across different scenarios. A researcher manually labeled these conversations as "Applicable" or "Not Applicable" under each of the five tags. To ensure validation, two other researchers independently labeled 20\% of the conversation data, checking for inter-rater reliability with Cohen's Kappa and Krippendorff's Alpha coefficients \cite{hayes2007answering}, both yielding values of 0.652, indicating substantial agreement among raters. 

We built Ordered Logistic Regression (OLR) models to predict user survey evaluation metrics based on five conversation tags, as summarized in Table \ref{tab:convo_olr}. A key finding was that the main factor contributing to user satisfaction with recommendations was the context provided by users. As shown in Fig. \ref{fig:convo_coding}, users who were dissatisfied with LLMRec typically shared less context compared to those who were satisfied and provided specific examples and requirements. The regression results also suggest that providing context helps LLMs generate more personalized and diverse recommendations. Additionally, context positively influences LLMRec's responsiveness, understandability, and general helpfulness. Meanwhile, \emph{Dialogue} had a negative impact on user-perceived understandability. This indicates that even when users communicated with the bot in plain, conversational language, they often felt that the bot didn't fully understand their objectives.

Users who tried to test LLMRec generally felt less satisfied with their interactions. However, since their goal was not to get movie recommendations, the negative impact of the \emph{Testing} label on recommendation or general interaction metrics is not necessary of concern. Additionally, there was a positive relationship between the \emph{Retry} label and recommendation explainability, indicating that repeating the same queries might improve understanding of why specific recommendations were made. However, Fig. \ref{fig:convo_coding} shows that repeating the same request is among the top patterns linked with negative interaction experiences, while using a chain-of-thought strategy could lead to higher satisfaction with the recommendations. With all the findings, we answer \emph{\textbf{RQ3}}:

    \emph{\textbf{RQ3}: What are some effective interaction strategies users can employ with the LLM to attain more satisfactory recommendations?}

Users need to provide adequate context and details about their preferences or requirements while asking LLM for recommendations. Providing examples of what they like or dislike is a good strategy to make LLM actually learn and expand upon. While having a human-like dialogue might not make the LLM understand user goals better, simply dumping queries and treating the LLMRec as a search engine without follow-up feedback and tuning is also not the optimal way to get good recommendations. Moreover, repeating the same queries might temporarily improve the explainability of recommendations, but that could not directly contribute to their ultimate satisfaction goals. In fact, some highly-satisfied users provided feedback or executed chain-of-thought strategy in their conversation. In addition, users need to be reminded about the ability boundary of LLM recommenders, that they should focus their conversation in the specific application domain, but not challenge the bot with irrelevant topics. 

    \begin{table*}[]
    \resizebox{2\columnwidth}{!}{%
    \begin{tabular}{@{}lllllllll@{}}
    \toprule
    \textit{\textbf{}} &
      \textbf{Satisfaction} &
      \textbf{Personalization} &
      \textbf{Diversity} &
      \textbf{Explainability} &
      \textbf{Responsiveness} &
      \textbf{Understandability} &
      \textbf{Helpfulness} &
      \textbf{Future Use} \\ \midrule
    \textit{Dialogue} & -.375(.086) & .031(.890)  & -.051(.821) & .040(.854)          & -.269(.240) & \textbf{-.427(.050)} & -.098(.665) & -.064(.775) \\
    \textit{Context} &
      \textbf{.464(.029)} &
      \textbf{.584(.007)} &
      \textbf{.748(.001)} &
      .210(.317) &
      \textbf{.564(.011)} &
      \textbf{.465(.027)} &
      \textbf{.513(.017)} &
      .188(.382) \\
    \textit{Steering} & .056(.793)  & -.127(.560) & .003(.991)  & .078(.715)          & -.124(.575) & -.281(.188)          & .162(.448)  & .186(.382)  \\
    \textit{Testing} &
      \textbf{-.938(.008)} &
      \textbf{-1.061(.010)} &
      -.816(.056) &
      {\color[HTML]{333333} \textbf{-.778(.028)}} &
      .301(.404) &
      -.645(.057) &
      -.456(.208) &
      \textbf{-1.026(.003)} \\
    \textit{Retry}    & -.021(.960) & .428(.299)  & .598(.119)  & \textbf{.845(.041)} & -.471(.212) & -.011(.977)          & .201(.608)  & .620(.118)  \\ \bottomrule
    \end{tabular}%
    }
    \caption{OLR Analysis of conversation label features to user survey response metrics. Each column of metrics is run with one single OLR model. Value in each cell is formatted as coef (p-val). }
    \label{tab:convo_olr}
    \vspace*{-0.8cm}
    \end{table*}

\vspace*{-0.2cm}
\section{Discussion}

In this section, we summarize our novel findings and future design implications of LLM recommenders from user experience perspectives in two folds: 1) How can RecSys researchers better utilize and improve current LLM modeling, prompting, or information retrieval strategies to better accommodate user needs; and 2) How can users be better informed or educated to achieve a more satisfactory interaction experience from LLM recommenders, without requiring specific technical background or understanding of terminology? 

\vspace{-0.2cm}
\subsection{Implications for LLM-Rec designers} 

Our findings showed that few-shot prompts didn't significantly improve recommendation quality compared to zero-shot or one-shot approaches. This aligns with \citeauthor{dai2023uncovering}, who noted that recommendation quality doesn't always improve with more examples~\cite{dai2023uncovering}. To address the lack of personalization and novelty in our findings, retrieval-augmented generation (RAG) has been proposed, allowing for user and item context to be integrated into LLM-based recommendations \cite{lewis2020retrieval}. Research by \citeauthor{di2023retrieval} has looked into retrieval-augmented RecSys with offline datasets \cite{di2023retrieval}, finding it effective for re-ranking more diverse, though potentially less relevant, recommendations \cite{carraro2024enhancing}. However, success with RAG requires careful selection of embedding and retrieval models, with field tests needed to validate effectiveness.

LLMs' native recommendation capabilities are often limited by their training data, which tends to be biased toward popular content \cite{zhang2023chatgpt}. However, LLMs offer two distinct advantages over traditional recommendation systems: 1) they excel at explaining recommendations using personal context, as noted by \citeauthor{acharya2023llm} \cite{acharya2023llm}, \citeauthor{wang2024understanding} \cite{wang2024understanding}, and \citeauthor{silva2024leveraging} \cite{silva2024leveraging}; 2) they can learn in-context, adapting to users' real-time interests during recommendation \cite{brown2020language,rubin2021learning}. We suggest that future practitioners leverage these benefits by integrating LLMs into recommendation systems to improve user experience. Additionally, a hybrid approach can help reduce LLM-generated content hallucination by anchoring recommendations in grounded, genuine item databases.

The qualitative analysis of survey responses revealed several areas for improving LLM-based recommenders. Users suggested incorporating beyond-text recommendations, like movie posters or trailers, for a more intuitive experience. This could be achieved with visual-enhanced chatbots or by exploring multimodal recommendation generation \cite{yuan2023go}. Users also wanted LLMRec to improve short-term memory and retain context throughout a conversation session, as seen in MemGPT \cite{packer2023memgpt}. Additionally, users requested LLMRec to be more proactive in seeking context and feedback during recommendations. Researchers might experiment with different LLM personas \cite{jiang2023personallm,mondal2024presentations} to determine the optimal level of proactivity and user-customizable recommendations.

Another issue raised was the balance between AI-generated content censorship and recommendation quality. Users proposed confidence thresholds and disclaimers to address LLM hallucinations and misinformation \cite{ferrara2024genai,luckett2023regulating,mesko2023imperative}. However, some users, like P241, expressed concerns about overly strict content guardrails: \emph{"...the guardrails were a bit too strict. I wanted to ask the system for a more naughty movie, but was prevented as it would not recommend anything with violence or sex. I can see why this is a problem in the US, but it is also a problem if the chatbot is unable to recommend high-quality movies just because they have a violent or sexual theme."} This highlights a research gap for future studies to explore the right balance of AI ethics in LLM recommenders.

\vspace{-0.2cm}
\subsection{Learnings for End Users} 

The success of LLM recommenders cannot be achieved without well-informed users with proper goals and interaction flows. The first learning we summarize from the study is to remind users to provide adequate context in order to get more personalized and specific recommendations that can better satisfy their needs, echoed by what \citeauthor{vaithilingam2022expectation} and \citeauthor{skjuve2023user} identified before \cite{vaithilingam2022expectation,skjuve2023user}. As we observed from the lack of difference in prompting techniques, it is challenging for designers to include the most relevant context to system prompt since users' needs are constantly changing. Future studies can explore different ways to encourage more context from users to benefit in-context learning \cite{brown2020language,rubin2021learning}. Some solutions can be tuning a proactive LLM that asks users for more contextual information. However, such model needs to be carefully tuned to respect the boundary between useful context and sensitive private information, and also keep conversation length in mind because users are impatient. Another way can be introducing some query examples for users to mimic or follow before they start their interaction.

Another important implication is to inform users about LLM recommender's capability and set reasonable expectations, similar to what \citeauthor{zamfirescu2023johnny} found before \cite{zamfirescu2023johnny}. As reflected in Fig. \ref{fig:convo_coding}, one of the top patterns associated with conversations of unsatisfied users it having too high expectation, such as P54 shared, \emph{"While the chatbot claims to know my profile, it clearly does not. It claimed that I recently disliked the Matrix, that is not true, I liked it, and it was a long time ago. But was right about the other movies."} This user case implies that while we need to set more strict rules for LLM to avoid such hallucination, we also need to help users construct the right expectation and understand there should be a room for LLM to make mistakes, which is also the case to encourage them to correct any wrong information and allow LLMs to learn. For example, users can be educated about how to utilize chain-of-thought interaction strategy to elicit step-wise reasons of making certain recommendations and iteratively improve the recommendations they get.

Finally, we also find future opportunity for developing built-in mechanism in LLM to detect users' off-the-track behavior and advise them to keep their conversation within the certain recommendation domain. Results from Fig. \ref{fig:convo_coding} and Table \ref{tab:convo_olr} both suggest that having irrelevant topics to recommendation or testing and challenging the LLM recommender contribute to negative user experience. To avoid unintentional drifting from the recommendation theme, users should be gently reminded of such behavior, either by LLM directly from conversation or from some extra UI alert.

\vspace{-0.2cm}
\subsection{Limitations}

Our study has several limitations. First, we were unable to test with more advanced pre-trained LLMs due to hardware constraints. Second, our prompts only incorporated recent user ratings, without exploring a broader range of user context combinations. Third, we only applied qualitative coding to user conversation data, without utilizing other NLP techniques for analysis. Addressing these limitations could be valuable for future research to enhance understanding in this domain.

\vspace*{-0.2cm}
\section{Conclusion}

In this study, we conducted a field experiment with LLM-based movie recommenders using zero-shot, one-shot, and few-shot personalized prompts to evaluate user interaction across three recommendation scenarios. By analyzing real-user conversations and semi-structured survey responses, we observed the following: 1) Users valued LLMs for their superior explainability and interactivity but found their generated recommendations not as good compared to those from classic recommender systems; 2) Recommendations for unpopular movies were more enjoyable and better met users' needs than those for personalized or ask-for-others scenarios. Additionally, the number of movies a user has watched significantly influenced their perception of LLM recommenders; 3) Providing personal or example-based context can lead to more personalized and satisfactory recommendations from LLMs. We believe these findings offer a useful starting point for future research aiming to enhance personalized interaction experiences with LLM-based recommender applications.

\bibliographystyle{ACM-Reference-Format}
\bibliography{ref}

\newpage
\appendix

\section*{Appendices} \label{appendix}

\setcounter{table}{0}
\renewcommand{\thetable}{A\arabic{table}}

    \begin{table}[H]
    \resizebox{\columnwidth}{!}{%
    \begin{tabular}{@{}llllll@{}}
    \toprule
    \textit{\textbf{}} & \textbf{F-val} & \textbf{P-val} & \textbf{F-O} & \textbf{F-Z} & \textbf{O-Z} \\ \midrule
    \textit{Enjoyment}         & .208  & .812 & .086(.825)  & .071(.867)  & -.015(.994) \\
    \textit{Satisfaction}      & 1.213 & .298 & -.197(.375) & -.191(.372) & .007(.999)  \\
    \textit{Personalization}   & 1.030 & .358 & .095(.791)  & -.113(.699) & -.208(.327) \\
    \textit{Diversity}         & 2.325 & .099 & -.235(.161) & -.232(.146) & .003(.100)  \\
    \textit{Novelty}           & 2.930 & .054 & -.266(.081) & -.238(.112) & .028(.973)  \\
    \textit{Serendipity}       & 2.002 & .136 & -.057(.894) & -.232(.132) & -.176(.342) \\
    \textit{Trustworthiness}   & .124  & .883 & -.057(.893) & -.048(.916) & .009(.997)  \\
    \textit{Explainability}    & .008  & .993 & -.008(.999) & -.018(.992) & -.001(.998) \\
    \textit{Responsiveness}    & 1.020 & .363 & -.161(.772) & -.321(.329) & -.161(.772) \\
    \textit{Understandability} & .278  & .758 & -.161(.765) & -.125(.839) & .036(.987)  \\
    \textit{Usefulness}        & .928  & .397 & -.140(.837) & -.321(.365) & -.182(.742) \\
    \textit{Future-Use}        & 1.107 & .333 & -.095(.935) & -.375(.325) & -.280(.559) \\ \bottomrule
    \end{tabular}%
    }
    \caption{AVOVA and pairwise Tukey's HSD test results for different prompt metrics. Pairwise stats include the mean difference and p-val in bracket. F: Few-shot, O: One-shot, Z: Zero-shot.}
    \label{tab:prompt-stats}
    \end{table}

    \begin{table*}[]
    \resizebox{2\columnwidth}{!}{%
    \begin{tabular}{@{}llllll@{}}
    \toprule
    \textbf{Example Raw User Conversation with LLMRec} &
      \textbf{Dialogue} &
      \textbf{Context} &
      \textbf{Steering} &
      \textbf{Testing} &
      \textbf{Retry} \\ \midrule
    "Can you recommend a few movies for me?" &
      NA &
      NA &
      NA &
      NA &
      NA \\
    \\
    \begin{tabular}[c]{@{}l@{}}"Can you recommend a movie for a date. it has to be something i like but she has to like it too. shes a bit more normal then me', \\ 'I dont want a rom com', 'something i havent seen', ':those are too weird for a first date with someone whos taste i dont know. \\ also ive seen 3 of those. try again. Ideally it would be a movie that maves the viewer feel sophisticated', '2 and 3 are good recs thanks.'\end{tabular} &
      A &
      A &
      A &
      NA &
      NA \\
    \\
    \begin{tabular}[c]{@{}l@{}}'Can you recommend some movie for me', 'i have watched inception', 'i have already watched the matrix', \\ 'Can you recommend some movie for me, 'but i have already watched inception', 'Can you recommend some movie for me', \\ 'but i have already watched inception, the shawshank redemption and the lords of the rings'\end{tabular} &
      NA &
      NA &
      A &
      NA &
      A \\
    \\
    \begin{tabular}[c]{@{}l@{}}"Ok, that seems useless and tedious, but let's try this. Recommend 10 movies", "I've seen all of those. Recommend 10 more movies \\ and don't recommend those again. Also, I hated The Godfather (1972) and The Big Lebowski (1998)", "I've seen all of those. \\ Recommend 10 more movies I haven't seen. I liked all of them except The Lobster (2015)", 'Who directed The Lighthouse (2019)?', \\ 'List all movies directed by Robert Eggers', "That isn't correct. Robert Eggers also directed The Northman and The Witch", \\ "I've watched Eternal Sunshine of the Spotless Mind (2004), The Lighthouse (2019), The Florida Project (2017), The Killing of a \\ Sacred Deer (2017), The Witch (2015), The Invitation (2015), The Love Witch (2016), and The Northman (2022). Recommend 10 more \\ movies I haven't seen. Don't recommend horror films.", 'Stop', "Recommend 10 movies I haven't seen that my boyfriend might also like", \\ 'I asked for 10 and you give me 5', "I can't find any information about The Prisoner (1977)"\end{tabular} &
      A &
      A &
      A &
      A &
      A \\
    \\
    \begin{tabular}[c]{@{}l@{}}'can you recommend some movies that are perfect for a road trip? preferably something with action, "recomend me a movie that I haven't seen", \\ 'something like "shutter island"', "can you recommend me  a good oscar worthy movie  preferably something I haven't seen", 'can you give me list', \\ "can you recommend me a list of good Thiller-worthy movies preferably something I haven't seen"\end{tabular} &
      NA &
      A &
      A &
      NA &
      NA \\ \bottomrule
    \end{tabular}%
    }
    \caption{Example user conversations and their associated taggings. A denotes Applicable, NA denotes Not Applicable. }
    \label{tab:example-user-convo-tagging}
    \end{table*}

\renewcommand{\thefigure}{A\arabic{figure}}
\setcounter{figure}{0}

    \begin{figure*}[]
        \centering
        \includegraphics[width=2\columnwidth]{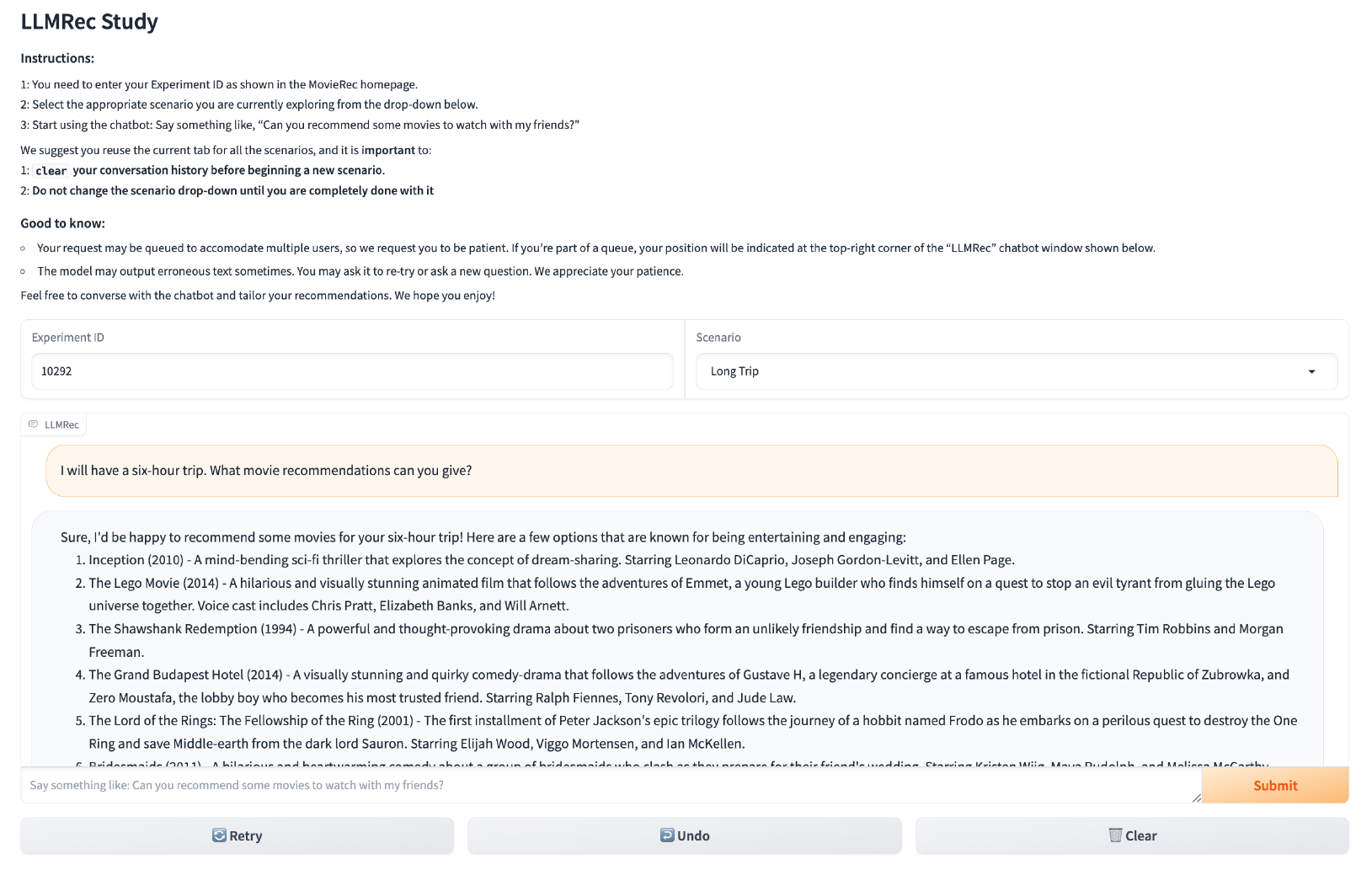}
        \caption{Example Chatbot Recommendation Interface.}
        \label{fig::example_chatbot_ui}
    \end{figure*}

\end{document}